\documentstyle[aps,epsf]{revtex}

\begin{document}
\draft\twocolumn \title{Quantum tomography of mesoscopic
superpositions of radiation states}
\author{G.M. D'Ariano, C. Macchiavello and L. Maccone} 
\address{
Theoretical Quantum Optics Group\\
Dipartimento di Fisica ``A. Volta'' and I.N.F.M., Via Bassi 6, 27100
Pavia, Italy}
\maketitle
\begin{abstract}
We show the feasibility of a tomographic reconstruction of
Schr\"{o}dinger cat states generated according to the scheme proposed
by S. Song, C.M. Caves and B. Yurke [Phys. Rev. A {\bf 41}, 5261
(1990)].  We present a technique that tolerates realistic values for
quantum efficiency at photodetectors. The measurement can be achieved
by a standard experimental setup.
\end{abstract}

\narrowtext

Optical homodyne tomography \cite{bilkent} is a powerful
tool to measure the density operator of a quantum system, with the
possibility of detecting purely quantum features of the radiation
field.  Such technique was suggested in the past to reconstruct the
density matrix of superpositions of classically distinguishable
quantum states \cite{mauro_qsp,DLP}, traditionally known as
Schr\"{o}dinger cats, which represent one of the most celebrated
examples of highly non classical states.  Some experiments have been
performed to detect Schr\"{o}dinger cat states in atomic systems
\cite{haroche}. For radiation, a scheme to detect cat states has been
proposed by Song, Caves and Yurke \cite{SCY}, however with no
feasibility study for a real experiment, concerning in particular the
main problem of quantum efficiency of detectors, which washes out the
fringes visibility.

In this paper we analyze a concrete experimental setup for the scheme
given in Ref. \cite{SCY}, and propose to detect the output field by
means of homodyne tomography.  We will examine the effects of quantum
efficiency and present a method to compensate them allowing a good
reconstruction of the Wigner function of the Schr\"{o}dinger cat, and
recovering the visibility of the experiment, even with quantum
efficiency $\eta_d=0.3$ at the readout photodetector, and $\eta_h=0.8$
at the homodyne detector.  Our proposal to reconstruct the Wigner
function allows to appreciate the whole anatomy of the cat rather than
just hearing its mew from the oscillations in a single quadrature
probability distribution.  To our knowledge, this is the first method
for detecting the density operator of Schr\"{o}dinger cat states for
free radiation, feasible with current technology.

Let us first briefly review the experimental scheme for generating
Schr\"{o}dinger cat states proposed in Ref.~\cite{SCY} (similar setups
were later proposed in Refs.~\cite{schleich,welsch}). The main idea
consists in feeding two orthogonally polarized modes of radiation,
called ``signal'' and ``readout'', both initially in the vacuum state,
into a parametric amplifier followed by a half-wave plate. The
parametric amplifier generates a correlated state of the two modes and
the half wave plate rotates the polarization directions by an angle
$\theta$.  The global state of the two modes at the output of this
setup is given by
\begin{equation}
|\psi\rangle=T(\theta)V(r)|0\rangle |0\rangle=
\sum_{j=0}^\infty \sum_{m=-j}^{j}B_{j,m}|j-m\rangle |j+m\rangle\;,
\label{psitot}
\end{equation}
where $V(r)=\exp[r(a_S a_R-a^\dagger _S a^\dagger_R)]$ describes the
action of the parametric amplifier, $T(\theta)=\exp[\theta(a_S
a^\dagger_R -a^\dagger _S a_R)]$ describes the polarization rotator and the 
coefficients $B_{j,m}$ are given by
\begin{eqnarray}
B_{j,m}=&&\frac{(-\tan\theta)^m(-\tanh r)^j}{\cosh r^{2}}
\sqrt{\frac{(j+m)!}{(j-m)!}}\nonumber\\
&&\sum_{k={max}(0,-m)}^j \frac{(j+k)!}{k!(j-k)!(m+k)!}(-\sin^2\theta)^k\;.
\label{B}
\end{eqnarray}
The rotation angle $\theta$ and the gain parameter $r$ are related by
the back-action-evading condition \cite{laporta,SCY} as $\sin
2\theta=\tanh r$.  The following step of the scheme consists in
detecting the number of photons at the readout mode. As a consequence
of this measurement, given $n_r$ photons detected at the readout, the
signal mode is reduced to the state
\begin{equation}
|\psi_{S,n_r}\rangle =
\frac{1}{P(n_r)}\sum_{j=0}^\infty B_{j+\lfloor\frac{n_r+1}{2}\rfloor,
j-\lfloor\frac{n_r}{2}\rfloor} |2j+\pi(n_r)\rangle\;,
\label{statorid}
\end{equation}
where $\lfloor x\rfloor$ denotes the integer part of $x$,
$\pi(n_r)$ is the parity of $n_r$ and
$P(k)$ is the probability of detecting $k$ readout photons with a perfect
photodetector, namely
\begin{equation}
P(k)=2^k\frac{(2k-1)!!}{(2k)!!}\frac{(\sinh r)^{2k}}{(2\sinh^2 r+1)^{k+1/2}}\;.
\label{pn}
\end{equation}
In the scheme of Ref. \cite{SCY}, after detection of the readout mode, 
the signal mode enters a
degenerate parametric amplifier with gain parameter $r_s$, 
described by the evolution operator
$S(r_s)=\exp[\frac{1}{2}r_s(a_S^2 - a_S^{\dagger 2})]$, which
increases the distance of the two components of the superposition in
the complex plane without changing the oscillating behavior of the
number probability distribution.  The final state of the signal is then
described by the following quadrature probability distribution
(the
quadrature operator is defined as $\hat x_\phi=\frac{1}{2}(a^\dagger e^{i\phi}
+a e^{-i\phi})$)
\begin{eqnarray}
&&P(X_S(\phi)|n_r)=\nonumber\\
&&\frac{(2{\mbox{Re}}\lambda/\pi)^{1/2}}
{(2n_r-1)!!\sigma^{n_r/2}}
e^{-2{\mbox{Re}}\lambda 
X_S^2(\phi)}\left |H_{n_r}(\sqrt{\lambda}X_S(\phi))\right |^2\;,
\end{eqnarray}
where the eigenvalue $X_S(\phi)$ of $\hat x_\phi$ 
is the field quadrature component at phase $\phi$, 
$H_{n}$ denotes the Hermite polynomial, 
$\lambda=[\cos\phi(e^{-2r_s} \cosh 2r\cos\phi+i\sin\phi)]^{-1}$
and $\sigma=1+\tan^2\phi\; e^{2r_s}/\cosh 2r$.

In this work we propose to detect the Schr\"{o}dinger cat at the signal
mode using optical homodyne tomography \cite{bilkent}. 
This technique consists in measuring the quadratures of radiation  
at several phases by means of a homodyne detector. The 
matrix elements of the density operator of the state are then measured as
follows
\begin{equation}
\langle \psi|\hat\rho |\psi'\rangle_{meas}=\overline{\langle \psi| 
{\cal K}_{\eta_h}(x-\hat x_\phi)|\psi'\rangle}\;,
\end{equation}
where ${\cal K}_{\eta_h}$ represents the kernel function given in
Refs. \cite{DLP,DMP} which depends on the value of the quantum
efficiency of the homodyne detector $\eta_h$, while the overbar
denotes the average over the experimental data at different
phases. The behavior of the kernel function sets the validity
limits of the tomographic reconstruction. These depend on the particular 
representation chosen to specify the density operator. In this work we always 
reconstruct the density matrix in the photon number representation: this is
possible for any value of the quantum efficiency $\eta_h > 1/2$ 
\cite{DLP,comment}. The $\eta_h=1/2$ bound for the overall homodyne
quantum efficiency is not a severe limitation in a real 
experiment, since good homodyne detectors can achieve values of 
$\eta_h$ between 0.85 and 0.94 \cite{schiller}.

\begin{figure}[hbt]
\vskip 1truecm
\begin{center}
\begin{center}\epsfxsize=.9 \hsize\leavevmode\epsffile{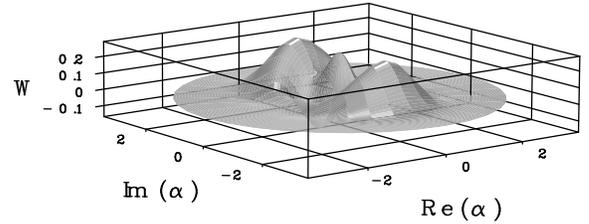}
\end{center}
\end{center}
\caption{Monte Carlo tomographic reconstruction of
the Wigner function of the state in Eq. (\ref{mixture}), with 
$\eta_d=0.3$, $\eta_h=0.8$, $r=r_s=0.4$ and $n_r=2$. Data are
collected for 70 different homodyne phases and $4\cdot 10^5$
 simulated data are used for each phase.}
\label{f:wnr}\end{figure}

\begin{figure}[hbt]
\vskip .3truecm\vspace{-1cm}
\begin{center}
\epsfxsize=.45\hsize\leavevmode\epsffile{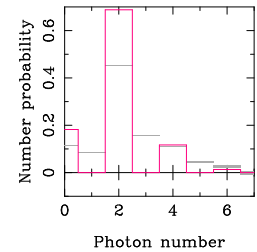}
\end{center}
\caption{Monte Carlo tomographic reconstruction of
the probability distribution for the same state and same simulated data 
of Fig. \ref{f:wnr}. The simulated values with the corresponding
statistical error bars are superimposed to the theoretical values
(solid line).}
\label{f:pnnr}\end{figure}

\begin{figure}[hbt]
\vskip .3truecm\vspace{-1cm}
\begin{center}
\epsfxsize=.45\hsize\leavevmode\epsffile{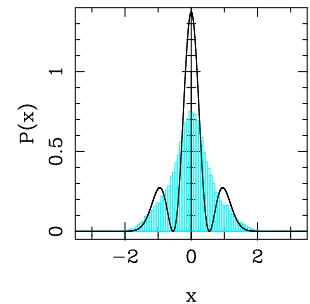}
\end{center}
\caption{Monte Carlo tomographic si\-mu\-la\-tion of
the qua\-dra\-ture pro\-ba\-bi\-li\-ty at $\phi=0$ for the same state of Fig. 
\ref{f:wnr}. The histogram contains 20000 simulated data, while the
solid curve is the theoretical distribution.}
\label{f:qdnr}\end{figure}

\begin{figure}[hbt]
\vskip .3truecm
\begin{center}
\epsfxsize=.8\hsize\leavevmode\epsffile{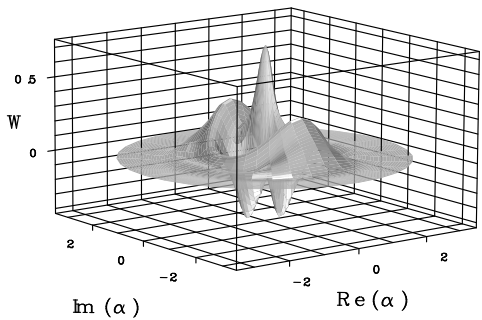}
\end{center}
\caption{Monte Carlo tomographic reconstruction of the Wigner function
of the state $|\tilde\psi_{S,2}\rangle$ with $\eta_d=0.3$,
$\eta_h=0.8$, $r=r_s=0.4$, using the reconstruction algorithm
(\ref{ric}). The same number of simulated data as in Fig. \ref{f:wnr}
is used.}
\label{f:rec}\end{figure}

Let us now consider the effect of non unit quantum efficiency $\eta_d$
at the readout photodetector. 
According to the Mandel-Kelley-Kleiner formula  \cite{eta}, a 
detector with non unit quantum efficiency is equivalent to 
a perfect photodetector preceded by a beam splitter with transmissivity 
$\eta_d$. Then, one can see that when $n_r$ photons are
detected at the readout, the signal mode is left in the following
statistical mixture of Schr\"{o}dinger cat states
\begin{eqnarray}
&&\hat\rho_{S,n_r}=\frac{1}{P_{\eta_d}(n_r)} \nonumber\\
&&\times\sum_{k=n_r}^\infty 
\left(\begin{array}{c} k \\ n_r \end{array}\right)
\eta_d^{n_r} (1-\eta_d)^{k-n_r} P(k) 
|\tilde\psi_{S,k}\rangle\langle\tilde\psi_{S,k}|\;,
\label{mixture}
\end{eqnarray}
where 
\begin{eqnarray}
|\tilde\psi_{S,k}\rangle=S(r_s)|\psi_{S,k}\rangle
\label{Spsi}
\end{eqnarray}
is the conditional Schr\"{o}dinger cat state at the signal mode
(\ref{statorid}) evolved by the degenerate parametric amplifier, and
$P_{\eta_d}(k)$ is the probability of detecting $k$ readout photons
with quantum efficiency $\eta_d$, i.e. the Bernoulli convolution of
the probability (\ref{pn}).  In Fig. \ref{f:wnr} we plot a Monte Carlo
tomographic reconstruction of the Wigner function of the statistical
mixture (\ref{mixture}) corresponding to an experiment with
$\eta_d=0.3$, $\eta_h=0.8$, $r=r_s=0.4$ and $n_r=2$. 
 As expected, the effect of non unit quantum
efficiency is to smooth the oscillations in the complex plane, which
are the typical signature of quantum interference (notice that the
theoretical Wigner function is practically indistinguishable from the
one plotted in Fig. \ref{f:rec}). Therefore, the
resulting state is more similar to a classical mixture of coherent
states rather than a Schr\"{o}dinger cat.  The degradation effects on
the cat due to non unit $\eta_d$ can be seen also in
Fig. \ref{f:pnnr}, where the number probability for the same
simulation is plotted: the probability still exhibits a non-monotonic
behavior, but the even terms no longer vanish. In Fig. \ref{f:qdnr}
we report a simulation of the quadrature probability distribution at
$\phi=0$, which would be seen following the original proposal
\cite{SCY}, with the corresponding theoretical curve.

We will now present a method to compensate these dramatic effects of realistic
quantum efficiencies. The main idea consists in the inversion of formula
(\ref{mixture}), which gives
\begin{eqnarray}
&&|\tilde\psi_{S,k}\rangle\langle\tilde\psi_{S,k}|=P(k)^{-1}
\eta_d^{-k}\nonumber\\
&&
\times\sum_{j=0}^\infty 
\left(\begin{array}{c} k+j \\ k \end{array}\right) (1-\frac{1}{\eta_d})^{j}
\hat\rho_{S,k+j}P_{\eta_d}(k+j)\;.
\label{ric}
\end{eqnarray}
Hence, a generic $k$-th Schr\"{o}dinger cat component of the signal mode can
be reconstructed by measuring all the signal states corresponding
to different readout numbers of photons (larger or equal to $k$), 
weighting each event 
according to Eq. (\ref{ric}).  In this way we have the additional advantage of using all data 
with $n_r\geq k$ than just those with
$n_r=k$ of the plain detection in Fig.  \ref{f:wnr}.
Moreover, by
processing the homodyne data according to Eq. (\ref{ric}) 
we can reconstruct the whole family of Schr\"{o}dinger cats 
$|\tilde\psi_{S,k}\rangle$ for different $k$'s at the same time. 
Notice that this method can be used for the reconstruction of any set of states
$|\psi_{S,k}\rangle$ conditioned by an inefficient photodetection.

In Fig. \ref{f:rec} we show a tomographic reconstruction of the same 
Schr\"{o}dinger cat component of Fig. \ref{f:wnr}, with the same 
values of the experimental parameters, but using the reconstruction procedure
based on the inversion (\ref{ric}). 
As we can see, all the oscillations in the Wigner function are
properly recovered, and the destructive effects of low quantum
efficiencies are defeated. 
Notice that unlike other compensation methods based on the inverse Bernoulli
transformation \cite{kiss}, where the convergence radius of the procedure
is $\eta>1/2$, the present method works also for very low values
of the quantum efficiency. In our case convergence of
the series (\ref{ric}) below the threshold $\eta_d=1/2$ is due to
the additional decaying factor $P_{\eta_d}(k+j)$.
An analysis of convergence of the series of errors as in Ref. \cite{comment}
shows that there is no lower-bound for $\eta_d$ if $r\leq \frac{1}{2}\ln(2+
\sqrt{3})\simeq 0.658$. Notice that the series convergence is slower for 
increasing $r$, which corresponds to more excited (macroscopic) 
Schr\"{o}dinger cats. This implies that the more macroscopic the cat is, 
the higher $\eta_d$ must be in order to have a good reconstruction.

\begin{figure}[hbt]
\vskip .3truecm
\begin{center}
\epsfxsize=.5\hsize\leavevmode\epsffile{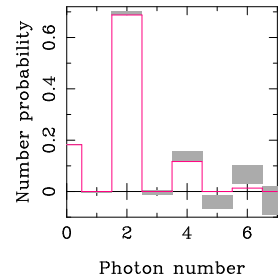}
\end{center}
\caption{Photon number probability of the cat
$|\tilde\psi_{S,2}\rangle$ using the reconstruction algorithm
(\ref{ric}), with the same parameters as in Fig. \ref{f:rec}. The
simulated values with the corresponding statistical error bars are
superimposed to the theoretical values (solid line).}
\label{f:num}\end{figure}

\begin{figure}[hbt]
\vskip .3truecm
\begin{center}
\epsfxsize=.45\hsize\leavevmode\epsffile{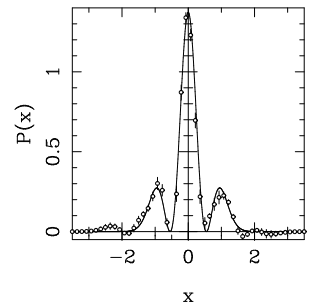}
\end{center}
\caption{Quadrature probability distribution at zero phase of the cat
state $|\tilde\psi_{S,2}\rangle$, with the same parameters as in
Fig. \ref{f:rec}. The simulated values with the corresponding
statistical error bars are superimposed to the theoretical curve
(solid line).}
\label{f:quad}\end{figure}

In Fig. \ref{f:num} we plot the number probability for the same parameters
of Fig. \ref{f:rec}: the simulated experimental results with corresponding
error bars are superimposed to the theoretical value.  As we can see, the 
tomographic reconstruction is very precise and the oscillations of the 
probability are perfectly resolved. In Fig. \ref{f:quad} we finally report
the quadrature probability distribution for $\phi=0$ superimposed to 
the theoretical curve. The visibility is totally recovered, in contrast to the
result in Fig. \ref{f:qdnr}, which would have been obtained according to the 
original proposal of Ref. \cite{SCY}.

Let us now discuss in more detail the experimental feasibility of the
scheme.  All the devices needed in the experiment are available with
the current technology. The parametric amplification can be realized
for example by an ordinary KPT crystal pumped with the second harmonic
of a mode locked Nd:Yag pulsed laser working at 80 MHz with 7 ps
pulses \cite{arecchi_pc}.  The major problem encountered to detect
Schr\"{o}dinger cats in conditional measurement schemes is the cat's
notorious fragility to any kind of losses and inefficiencies.  The
novelty of the present proposal is that using the reconstruction
method based on Eq. (\ref{ric}) low values of $\eta_d$ can be
tolerated, and hence ordinary linear avalanche photodiodes with
$\eta_d\sim$0.3 can be used.  On the other hand, the tomographic
apparatus needed to detect the Schr\"{o}dinger cat at the signal mode
is based on homodyne detection, with the possibility of using
high--efficiency photodetectors, because single-photon resolution is
no longer needed due to the amplification from the local oscillator
(LO). Moreover, the LO comes from the same laser source of the
classical pump of the OPA, in order to achieve time matching of
modes. In addition, since there is no fluctuating phase in the whole
optical setup, neither in the second harmonic generation stage nor in
the homodyne detection, the LO is also perfectly phase matched with
the pump.  The resulting setup is very stable and can take
measurements for tens of minutes at a rate of $10^8$ data/sec at the
readout photodetector. The tomographic reconstructions presented in
this paper were obtained with $2.8\cdot 10^7$ experimental data. In
these examples the probability of detecting less than two photons at
the readout photodetector is $\sim 0.9967$.  Therefore, taking into
account that only the fraction $3.3\cdot 10^{-3}$ of experimental data
collected at the readout is useful for the Schr\"{o}dinger cat
reconstruction, we can easily see that the whole set of data can be
collected in a few minutes.  Notice that increasing $k$, which
corresponds to a more excited cat, the number of useful data
decreases, and the reconstruction becomes slower. For example, to
reconstruct the cat component with $k=4$ only $2\cdot 10^{-5}$ of the
experimental data are useful, for $k=5$ we can use only the fraction
$1.7\cdot 10^{-6}$ of data, and so on. Moreover, to reconstruct more
excited cat components, higher index density matrix elements are
needed for the Wigner function, and the effect of statistical errors
from tomography becomes more dramatic \cite{errori}, with the
consequence that more data are needed to reach a prescribed
accuracy. For these reasons, the more excited the cat is, the longer
the experiment and the more difficult the state is to detect. Finally,
it has been suggested \cite{ar-mon} that by lowering the gain $r$ some
(non tomographic) homodyne visibility could be detected for high
$\eta_h$ ($>0.9$) and for $\eta_d$ as low as $0.3$, even without our 
reconstruction algorithm. However, by lowering the gain the data
acquisition rate is greatly reduced, whereas our method works also
with high gains.

In conclusion, we have shown the feasibility of a tomographic
reconstruction of a Schr\"{o}dinger cat in an experimental scheme
which is practical in a laboratory using standard technology devices.
The problem of low efficiencies at the single-photon resolving
detector, which was regarded as the major obstacle for experiments of
this kind \cite{SCY}, has been solved by the implementation of a
suitable tomographic data processing. The whole density matrix of the
cat, and hence all its characteristics (such as the photon number
probability, the quadrature distribution and the Wigner function), can
be measured in this way, whereas the plain homodyne detection proposed
in the original scheme of Ref. \cite{SCY} would not have provided
visible probability oscillations with the available low-efficiency
single-photon-resolving detectors.

We thank T.F. Arecchi for illuminating discussions on the experimental
setup. This work has been supported by the PRA--CAT97 of the INFM.

\end{document}